\documentclass[graybox]{svmult}

\usepackage{mathptmx}       
\usepackage{helvet}         
\usepackage{courier}        
\usepackage{makeidx}         
\usepackage{graphicx}        
\usepackage{multicol}        
\usepackage[bottom]{footmisc}

\makeindex             

\begin{document}

\title*{Benchmarking ordering techniques for nonserial dynamic programming\thanks{This research is partly supported by FWF (Austrian Science Funds) under the project P20900-N13.}
}
\author{Alexander Sviridenko and Oleg Shcherbina}
\institute{A. Sviridenko \at
              Faculty of Mathematics and Computer Science \\
              Tavrian National University, Vernadsky Av. 4,  Simferopol 95007  Ukraine\\
              \email{oleks.sviridenko@gmail.com} 
\and
           O. Shcherbina \at
              Faculty of Mathematics, University of Vienna\\
              Nordbergstrasse 15,  A-1090 Vienna,  Austria\\
              \email{oleg.shcherbina@univie.ac.at}
}

%
%
\maketitle

\abstract{Five ordering algorithms for the nonserial dynamic programming
algorithm for solving sparse discrete optimization problems are
compared in this paper. The benchmarking reveals that  the
ordering of the variables has a significant impact on the
run-time of these algorithms. In addition, it is shown that
different orderings are most effective for different classes
of problems. Finally, it is shown that, amongst the algorithms
considered here, heuristics based on maximum cardinality search
and minimum fill-in perform best for solving the discrete
optimization problems considered in this paper.}
\section{Introduction}
\label{intro}
The use of discrete optimization (DO) models and algorithms
makes it possible to solve many real-life problems in scheduling
theory, optimization on networks, routing in communication networks,
facility location in enterprize resource planing, and logistics.
Applications of DO in the artificial intelligence field include theorem
proving, SAT in propositional logic, robotics problems, inference calculation
in Bayesian networks, scheduling, and others.\\
Many real-life discrete optimization problems (DOPs) contain a huge number
of variables and/or constraints that make the models intractable for currently
available DO solvers. Usually, such problems have a special structure, and the
matrices of constraints for large-scale problems are sparse. The nonzero elements
of the matrices often involve a limited number of blocks. The block form of
many DO problems is usually caused by the weak connectedness of subsystems of
real-world systems.\\
One of the promising ways to exploit sparsity for solving sparse DOPs is
nonserial dynamic programming (NSDP) \cite{BerBri}, \cite{Soa07}.
NSDP eliminates variables of DOP using an elimination order which makes
significant impact on running time. As finding an optimal ordering is
NP-complete \cite{Yanna81}, heuristics are utilized in practice for finding
elimination orderings. The literature has reported extensive computational
results for the use of different ordering heuristics in the solution of systems
of equations \cite{Amestoy96}, \cite{George73}.  However, no such experiments
have been reported for NSDP to date.\\
Given the increased recent interest in DOPs, the subject of experimental
research of NSDP algorithms that utilize heuristic variable orderings is
timely. In this paper, we present  comparative computational results from
the benchmarking of five ordering techniques, namely: minimum degree ordering,
nested dissection ordering, maximum cardinality search, minimum fill-in, and
lexicographic breadth-first search.

\section{Nonserial Dynamic Programming Algorithm}
\label{sec:1}
Consider a DOP with constraints:
\begin{equation}\label{gen1}
 F(x_{1}, x_{2}, \ldots, x_{n})=\sum_{k \in K} f_{k}(Y^{k}) \rightarrow \max
\end{equation}
subject to the constraints\\
\begin{equation}\label{gen2}
 g_{i}(X_{S_i})~ R_{i}~ 0, ~~i \in M = \lbrace 1, 2, \ldots, m \rbrace,
 \end{equation}
 \begin{equation}\label{gen3}
x_{j} \in D_{j}, ~~j \in N= \lbrace 1,\ldots, n \rbrace,
\end{equation}
where\\
$X = \left\{ x_1,\dots,x_n \right\}$ is a set of discrete
variables, $Y^{k} \subseteq \lbrace x_{1}, x_{2}, \ldots, x_{n}\rbrace,
k \in K=\left\{1,2,\ldots,t \right\},$ $t$ -- number of components of
objective function, $S_{i}\subseteq \{ 1,2, \ldots, n\},~  R_{i} \in
\lbrace \leq, = ,\geq \rbrace, i \in M$; $D_j$ is a finite set of admissible
values of variable $x_j,~~j \in N$.
 The functions $f_k(X^k),~~k \in K$ are called components of the objective function
and can be defined in tabular form. We use here a notation: if $S=\{j_1,\ldots,j_q\}$
then $X_S=\{x_{j_1},\ldots,x_{j_q}\}$.\\
\subsection{The structure of sparse discrete optimization problems}
\label{sec:11}
Let us take a detailed look at an NSDP implementation for solving
DO problems for the case when the structural graph is an interaction
graph of variables.

\begin{definition} \cite{BerBri}. Variables  $x \in X$ and $y \in X$
interact in  DOP with constraints (we denote  $x \sim y$) if they appear
both either in the same component of objective function, or in the same
constraint (in other words, if variables are both either in a set  $X^k$,
or in a set  $X_{S_i}$).
\end{definition}
\begin{definition} \cite{BerBri}. Interaction graph of the DOP is called
an undirected graph  $G=(X, E)$, such that
\begin{enumerate}
\item Vertices $X$ of $G$ correspond to variables of the DOP;
\item Two vertices of $G$ are adjacent if corresponding variables interact.
\end{enumerate}
\end{definition}
Further, we will use the notion of vertices that correspond one-to-one to variables.

\begin{definition} Set of variables interacting with a variable  $x \in X$,
is denoted as  $Nb(x)$ and called a neighborhood of the variable   $x$.
For corresponding vertices of $G$ a neighborhood of a vertex   $x$ is a set
of vertices of interaction graph that are linked by edges with   $x$.
Denote the latter neighborhood as  $Nb_G(x)$.
\end{definition}
In hypergraph representation of DO problems structure, the set of vertices $H$
of hypergraph, equals to the set of variables  $X$  from the DO problems, and
hypergraph's hyperedges forms subsets of related variables that are included
in constraints, which means the hyperedge defines constraint scope.

\subsection{NSDP (Variable elimination) algorithms}
\label{sec:12}
Consider a sparse discrete optimization problem (\ref{gen1}) -- (\ref{gen3}) whose
structure is described by an undirected interaction graph  $G=(X, E)$. Solve this problem with a NSDP.\\
Given ordering  of variable indices $\alpha$  the NSDP proceeds in the following way:
it subsequently eliminates $x_{\alpha_1},\dots, x_{\alpha_n}$  in the current graph and
computes an associated local information $h_i(Nb(x_{\alpha_i}))$ about vertices from
$Nb(x_{\alpha_i})$ ($i=1,\dots,n$). This can be described by the combinatorial elimination process:
\[
G^0=G,G^1, \ldots, G^{j-1}, G^{j}, \ldots, G^{n},
\]
where  $G^{j}$ is the  $x_{\alpha_j}$-elimination graph of $G^{j-1}$ and $G^n = \emptyset$\\
The process on interaction graph transformation corresponding to
the NSDP scheme is known as elimination game which was first introduced
by Parter \cite{Part} as a graph analogy of Gaussian elimination.
The input of the elimination game is a graph $G$ and an ordering $\alpha$ of $G$. \\
Consider the DOP described above and suppose without loss of generality that
variables are eliminated in the order  $x_1,\dots, x_n$ .
Using the variable elimination scheme eliminate a first variable  $x_1$.
This  $x_1$  is in a set of constraints with the indices of  $U_1=\{i~|~x_1 \in X_{S_i}\}$.
Together with   $x_1$, in constraints  $U_1$ are variables from  $Nb(x_1)$.
To the variable $x_1$  corresponds the following subproblem  $P_1$:
\[
h_{1}(Nb(x_{1}))=\max_{x_1} \Bigl\{\sum_{k \in K_1} f_{k}(Y^{k})~|~g_{i}(X_{S_i})~ R_{i}~ 0, ~~i \in  U_{1},~ x_{j} \in D_j,~ x_{j} \in Nb[x_{1}] \Bigr\}.
\]
Then the initial DOP can be transformed in the following way:
\[
\max_{x_1,\ldots, x_n} \Bigl\{\sum_{k \in K} f_{k}(Y^{k})~|~g_{i}(X_{S_i})~ R_{i}~ 0, ~~i \in  M,~ x_{j} \in D_j,~ j \in N \Bigr\}=
\]
\[
\max_{x_2,\ldots, x_{n}} \Bigl\{ \sum_{k \in K-K_1} f_{k}(Y^{k}) +
h_{1}(Nb(x_{1})~|~ g_{i}(X_{S_i})~ R_{i}~ 0, ~~i \in  M-U_{1},~ x_{j} \in D_j,~ j \in N - \{1\}\Bigr\}
\]
The last problem has  $n-1$ variables; from the initial DOP were excluded constraints 
with the indices from $U_{1}$  and objective function term  $\sum_{k \in K_1} f_{k}(Y^{k})$; 
there appeared a new objective function term $h_{1}(Nb(x_{1}))$. Due to this fact 
the interaction graph associated with the new problem is changed: a vertex $x_{1}$  
is eliminated and its neighbors have become connected.\\
Denote the new interaction graph $G^1$  and find all neighborhoods of variables in  $G^1$. 
NSDP eliminated the remaining variables one by one in an analogous manner.

\section{Elimination Ordering Techniques}
\label{sec:2}
An efficiency of the NSDP algorithm crucially depends on the interaction graph 
structure of a DOP. If the interaction graph is rather sparse or, in other words, 
has a relatively small induced width, then the complexity of the algorithm is reasonable. 
At the same time an interaction graph leads us to another critical factor such as an 
elimination order which should be obtained from the interaction graph.\\
From the other side the NSDP algorithm heavily depends on the elimination ordering. 
A good elimination ordering yields small cliques during variable elimination. There 
are several successful schemes for finding a good ordering which we will used in this 
paper: \textit{minimum degree ordering algorithm} (MD), \textit{nested dissection ordering algorithm} 
(ND), \textit{maximum cardinality search algorithm} (MCS), \textit{minimum fill-in heuristic} 
(MIN-FILL) and \textit{lexicographic breath-first search algorithm} (LEX-BFS).
\subsection{ Minimum degree ordering algorithm}
\label{sec:21}
The minimum degree (MD) ordering algorithm \cite{Amestoy96} is one of the 
most widely used in linear algebra heuristic, since it produces factors 
with relatively low fill-in on a wide range of matrices.\\
In the minimum degree heuristic, a vertex $v$  of minimum degree is chosen. 
The graph  $G'$, obtained by making the neighborhood of $v$  a clique and then 
removing  $v$ and its incident edges, is built. Recursively, a chordal supergraph  
$H'$ of   $G'$ is made with the heuristic. Then a chordal supergraph $H$  of  $G$ 
is obtained, by adding   $v$ and its incident edges from  $G$ to  $H'$.
To create an elimination order with help of minimum degree ordering algorithm 
the \textbf{$minimum\_degree\_ordering()$} function from BOOST library \cite{BOOST} has been used.
\subsection{ Nested dissection algorithm}
\label{sec:22}
To create an elimination order, we recursively partition the elimination graph 
using nested dissection. More specifically, we use \textbf{$METIS\_EdgeND()$} function 
from METIS library \cite{Karypis} to find a nested dissection ordering.
\subsection{Maximum cardinality search algorithm}
\label{sec:23}
The Maximum Cardinality Search (MCS) algorithm \cite{Tarjan84} visits the vertices of a
graph in an order such that at any point, a vertex is visited that has the
largest number of visited neighbors. An MCS-ordering of a graph is an ordering 
of the vertices that can be generated by the Maximum Cardinality Search algorithm. 
The visited degree of a vertex $v$  in an MCS-ordering is the number of neighbors of  $v$ 
that are before  $v$ in the ordering.\\
To create an elimination ordering the \textbf{$chompack.maxchardsearch()$} function from the
\textbf{Chordal Matrix Package} (CHOMPACK) \cite{CHOMPACK} has been used.
\subsection{Minimum Fill-in algorithm}
\label{sec:24}
The minimum fill-in heuristic \cite{Jegou} works similarly with minimum degree 
heuristic, but now the vertex $v$ is selected such that the number of edges that 
is added to make a neighborhood of  $v$   a clique is as small as possible.
\subsection{ Lexicographic breadth-first search algorithm}
\label{sec:25}
Lexicographic breadth-first search algorithm (LEX-BFS) \cite{Rose76}  numbers the vertices
from $n$  to 1 in the order that they are selected.
This numbering fixes the positions of an elimination scheme. For each vertex  $v$,
the label of $v$  will consist of a set of numbers listed in decreasing order. 
The vertices can then be lexicographically ordered according to their labels.
\section{Benchmarking}
\label{sec:3}
\subsection{ NSDP algorithm implementation}
\label{sec:31}
The NSDP algorithm was implemented by the first author in Python. 
The ND and MD algorithms were implemented in C and C++, respectively. 
To work with graph objects was taken class \textbf{networkx.Graph} from 
the \textbf{networkx} library \cite{NetworkX}.
\subsection{ Test problems}
\label{sec:32}
For benchmarking the DO test problems were generated by using hypergraphs 
from the CSP\footnote{CSP -- Constraint Satisfaction Problem.}  hypergraph 
library \cite{Musliu}. This collection contains
various classes of constraint hypergraphs from industry (DaimlerChrysler,
NASA, ISCAS) as well as synthetically generated ones (e.g. Grid or Cliques).\\
The test problems were generated in the following way. The constraints structure
of a linear DO problem with binary variables was described by hypergraph from
the library \cite{Musliu}. To build constraint  $i$  the next hyperedge of hypergraph
was taken, which includes a set of variables $X_{S_i}$  for a new building constraint.
In the next step, the coefficients for appropriate variables of $A_{S_i}$ were generated
using a random number generator. Then the left part of  $i$-th constraint had view
$A_{S_i}X_{S_i}$, while the right part was  $\sigma \sum A_{S_i}$, where  $\sigma$
is random number from interval (0, 1). Objective function is linear and includes all
variables -- vertices of hypergraph, where coefficients $c_j$  of objective function
$\sum_{j=1}^n c_j x_j \rightarrow \max$ where created with help of random number generator.\\
After the test problems were generated, the ordering algorithms MD, ND, MCS, MIN-FILL 
and LEX-BFS were applied for obtaining an elimination ordering. 
Then the problems were solved with the NSDP algorithm 
by utilizing to the specified elimination ordering.
\subsection{ Benchmarking ordering analysis}
\label{sec:33}
\begin{figure*}
  \includegraphics[width=0.75\textwidth]{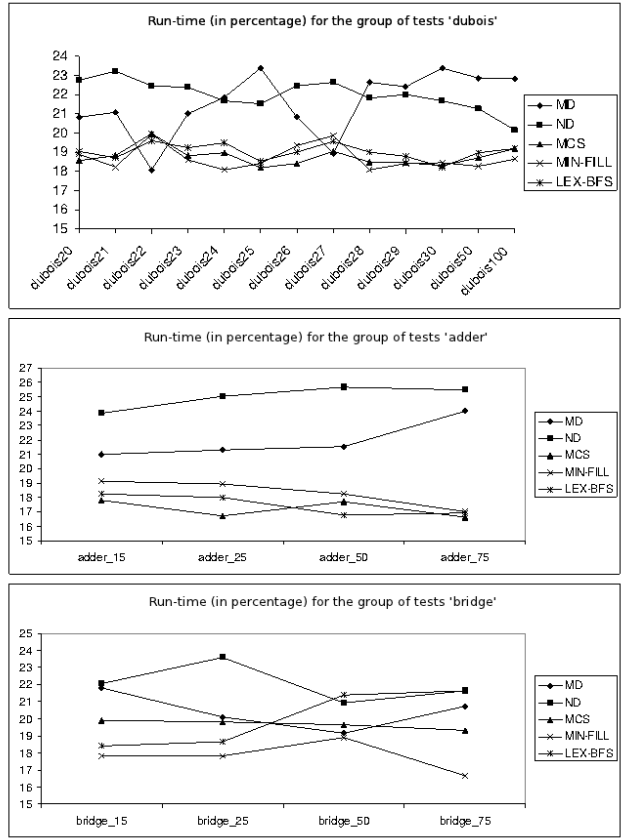}
\caption{Run-time (in seconds) for groups of tests 'dubois', 'adder', 'bridge'.}
\label{fig:1}       
\end{figure*}

\begin{figure*}[h!]
  \includegraphics[width=0.75\textwidth]{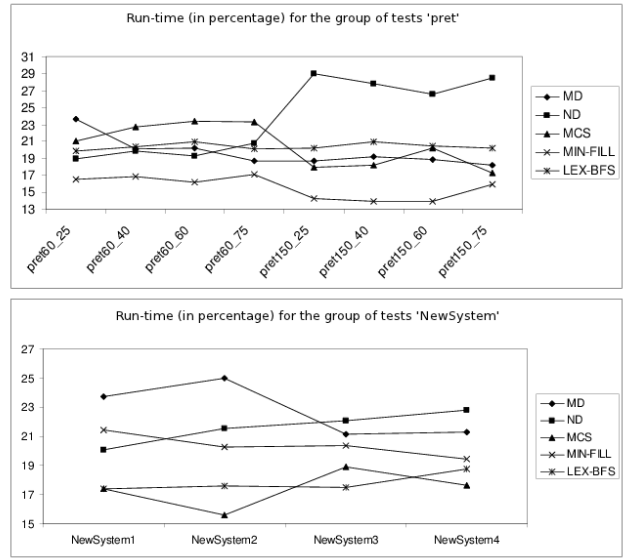}
\caption{Run-time (in seconds) for groups of tests 'pret' and 'NewSystem'.}
\label{fig:2}       
\end{figure*}

The following five groups of 33 test problems have been taken: 'dubois', 
'bridge', 'adder', 'pret' and 'NewSystem'. All experimental results were 
obtained on a machine with Intel Core 2 Duo
processor \@ 2.66 GHz, 2 GB main memory and operating system Linux, 
version 2.6.35-24-generic. The results can be found
in table 1, in which $n$ denotes the number of variables, $m$ the  number of constraints and the minimal time of
problem solving for appropriate heuristics was underlined.  We can see 
that for ND algorithm the minimal run-time of the NSDP algorithm was 
achieved 0 times (0 \%), for MD 2 times (6,0 \%), LEX-BFS 3 times (9,1 \%), 
MCS 9 times (27,3 \%) and MIN-FILL 19 times (57,6 \%).\\
\begin{table}[h!]
\caption{Run-time (in seconds) of solving DO problems with help of NSDP algorithm using different orderings.}
\label{tab:1}       
\begin{tabular}{lrrrrrrr}
\hline\noalign{\smallskip}
Test  &  $n$  & $m$ &  MD & ND & MCS & MIN-FILL  &  LEX-BFS\\
\hline\noalign{\smallskip}
dubois20&60&40&1,31&1,43&\underline{1,17}&1,19&1,20\\
dubois21&63&42&1,52&1,67&1,35&\underline{1,31}&1,35\\
dubois22&66&44&\underline{1,37}&1,70&1,51&1,51&1,49\\
dubois23&69&46&1,90&2,02&1,70&\underline{1,68}&1,74\\
dubois24&72&48&2,18&2,17&1,89&\underline{1,80}&1,95\\
dubois25&75&50&2,72&2,50&\underline{2,11}&2,14&2,15\\
dubois26&78&52&2,62&2,82&\underline{2,32}&2,43&2,39\\
dubois27&81&54&\underline{2,58}&3,09&2,60&2,71&2,66\\
dubois28&84&56&3,55&3,43&2,90&\underline{2,84}&2,98\\
dubois29&87&58&3,91&3,84&3,22&\underline{3,21}&3,28\\
dubois30&90&60&4,46&4,14&3,50&3,52&\underline{3,48}\\
dubois50&150&100&17,52&16,31&14,34&\underline{14,00}&14,53\\
dubois100&300&200&126,32&111,64&106,34&\underline{103,23}&106,29\\
adder\_15&106&75&6,20&7,04&\underline{5,25}&5,64&5,38\\
adder\_25&176&125&27,33&32,13&\underline{21,47}&24,27&23,10\\
adder\_50&351&250&326,54&388,74&268,39&276,47&\underline{254,25}\\
adder\_75&526&375&4876,74&5180,46&\underline{3381,28}&3460,32&3435,76\\
bridge\_15&137&135&15,21&15,41&13,90&\underline{12,45}&12,86\\
bridge\_25&227&225&63,14&74,18&62,30&\underline{55,98}&58,62\\
bridge\_50&452&450&900,20&983,40&922,68&\underline{886,91}&1003,89\\
bridge\_75&677&675&4832,55&5049,61&4507,38&\underline{3886,34}&5040,91\\
pret60\_25&60&53&1,89&1,51&1,68&\underline{1,32}&1,59\\
pret60\_40&60&53&1,50&1,48&1,69&\underline{1,26}&1,52\\
pret60\_60&60&53&1,58&1,51&1,83&\underline{1,27}&1,64\\
pret60\_75&60&53&1,44&1,60&1,79&\underline{1,31}&1,54\\
pret150\_25&150&133&21,29&33,06&20,51&\underline{16,25}&23,04\\
pret150\_40&150&133&22,63&32,83&21,47&\underline{16,44}&24,76\\
pret150\_60&150&133&22,09&31,13&23,72&\underline{16,35}&23,99\\
pret150\_75&150&133&21,20&33,23&20,19&\underline{18,57}&23,59\\
NewSystem1&142&85&19,21&16,24&\underline{14,05}&17,33&14,08\\
NewSystem2&345&200&603,83&520,82&\underline{376,33}&489,80&425,35\\
NewSystem3&474&284&1294,86&1352,75&1159,75&1247,81&\underline{1072,58}\\
NewSystem4&718&422&7769,10&8322,58&\underline{6427,85}&7095,17&6845,20\\
&&&&&&&\\
\hline\noalign{\smallskip}
\end{tabular}
\end{table}
Let us take a look at the benchmarking results in more detail. Fig. \ref{fig:1} describes
results of the experiment for the groups of test problems 'dubois', 'bridge', and
'adder'. These results show to us that MCS, MIN-FILL and LEX-BFS heuristics behave 
quite similar and give the best result for a given group of problems.
At the same time MD and ND show the worst time result. Also for the group 'bridge' 
we can see the gradual decreasing of the time result for the LEX-BFS heuristics and the obvious domination of MIN-FILL.\\
Fig. \ref{fig:2} describes experimental results for the groups of test problems
'pret' and 'NewSystem'.
Here we can see the importance of the right choice
of heuristics for a certain group of problems. In the case of 'pret', we 
see obvious domination of MIN-FILL algorithm, while MCS and LEX-BFS fall behind. However, the group 'NewSystem' shows completely opposite results, where MIN-FILL runs third while MCS and LEX-BFS take the first two places.\\
In the case of 'pret', we see the obvious domination of the MIN-FILL algorithm, while MCS and LEX-BFS go back. But the group 'NewSystem' shows the completely opposite result, where MIN-FILL goes on the third place while MCS and LEX-BFS take the first places.
\section{Conclusion}
The goal of this paper to research the role of five variable ordering 
algorithms and to describe the effect that they play on solving time of 
sparse DO problems with help of NSDP algorithm. Our computational experiments 
demonstrate that, for solving DO problems, variable ordering has a significant 
impact on the run-time for solving the problem. Furthermore, different ordering 
heuristics were observed to be more effective for different classes of problems. 
Overall, the MCS and MIN-FILL heuristics have provided the best results for solving 
DO problems of the problem classes that were considered in this paper.
It seems promising to continue this line of research by studying methods of 
block elimination with suitable partitioning methods.

%

%





\end{document}